\documentclass[twocolumn,aps,prd,superscriptaddress,nofootinbib]{revtex4-2}
\usepackage{graphicx}
\usepackage{subfig}
\usepackage{color}
\usepackage{amsmath}
\usepackage{amssymb}
\usepackage{bm}
\usepackage{slashed}
\usepackage{epsfig}
\usepackage{amsfonts}
\usepackage{epstopdf}
\usepackage{textcomp}
\usepackage{enumitem}
\usepackage{subfig}
\usepackage{hyperref}
\usepackage{cancel}
\usepackage{appendix}
\usepackage{ amsmath }
\usepackage{diagbox}
\usepackage{dcolumn}% Align table columns on decimal point
\usepackage{lipsum} % 
\begin{document}
\title{A new approach for amplitudes with multiple fermion lines}	

	\author{Feng Zhang}
	\email{zhangfeng@ihep.ac.cn} 
	\author{Bin Gong}
	\email{twain@ihep.ac.cn} 
	\author{Jian-Xiong Wang}
	\email{jxwang@ihep.ac.cn}
	\affiliation{
		Institute of High Energy Physics (IHEP), Chinese Academy of Sciences (CAS),
		19B Yuquan Road, Shijingshan District, Beijing, 100049, China}
	\affiliation{University of Chinese Academy of Sciences (UCAS), Chinese Academy of Sciences (CAS),
		19A Yuquan Road, Shijingshan District, Beijing, 100049, China}

\date{\today}

\begin{abstract}
%{\color{red}
A new approach for tree-level amplitudes with multiple fermion lines is presented. 
It mainly focuses on the simplification of fermion lines.
By calculating two vectors recursively without any matrix multiplications, the result of a fermion line is reduced to a very compact form depending only on the two vectors. 
The comparisons with other packages are presented, and the results show that our package FDC gives a very good performance in the processes 
of multiple fermion lines with this new approach and some other improvements. A further comparison with WHIZARD shows that this new approach 
has a competitive efficiency in computing pure amplitude square without phase space integration.

\end{abstract}

\maketitle
%\allowdisplaybreaks

\section{Introduction}
As the development of high energy physics, the sensitivity of detectors has been improved and a large number of data samples is accumulated, which makes the experimental measurements more and more precise.
Therefore, the predictions from the theoretical side need to be precise enough to match the measurements, which demands
the calculation of higher order perturbative corrections.
Fortunately, the technology of Feynman amplitude calculation has also been greatly improved in recent years. Nowadays automatic one-loop calculation is already available, provided by many different packages (see e.g.~\cite{Hahn:2016ebn, Alwall:2014hca, Kilian:2007gr, Cullen:2014yla, Belanger:2003sd}). Automatic multi-loop calculation is still under process, but there are already many progresses~(see .e.g.~\cite{Borowka:2015mxa,Borowka:2017idc,Smirnov:2015mct,Smirnov:2019qkx,Liu:2022chg}).

Meanwhile, multi-particle processes become more and more important in this procedure.
For example, it is well-known that  in future $e^{+}e^{-}$ colliders such as the International Linear Collider (ILC) and the Compact Linear Collider (CLIC) the pair production of top quark $e^{+}e^{-}\rightarrow t\overline{t}$ is very important~\cite{Horiguchi:2013wra,Amjad:2015mma,Seidel:2013sqa,Abramowicz:2018rjq,Fuster:2015jva}. In this process, top pair will decay which results a series of subprocesses with 6 final states. 
%, which requires not only that we can calculate this process with so many particles, but also that the calculation speed should be fast.
%%}
On one hand due to the high luminosity such subprocesses can be directly measured and studied. 
On the other hand their contributions are comparable with certain higher order corrections thus should not be neglected even in perturbative calculation.

The calculation of such processes is straightforward but cumbersome, even at tree level. It's because the number of Feynman diagrams grows very rapidly as external particles increases such that the expression of total scattering amplitude becomes very complicated. Despite of some special cases, such calculation can only be done numerically.

Things becomes worse if there exists fermions in external particles. The scattering amplitude of a process with external fermions can roughly be expressed as
\begin{equation}
{\cal M} \sim \sum\limits_{i} F_{i_1}F_{i_2}\cdots F_{i_n}.
\label{ee1}
\end{equation}
We have used $\sim$ in the expression since the coefficients have been dropped. Here $F_k$ denotes a fermion line, which contains  a string of Dirac $\gamma$-matrices and two spinors, i.e.
\begin{equation}
F_k = \overline{U}_1 \widehat{V}_1\cdots \widehat{V}_m U_2.
\end{equation}
Here $U_{1,2}$ are the two spinors, and $V_{1,2,\cdots,m}$ are several vectors, which can be external momenta, polarization vectors and their combinations.
We use a hat ( $\widehat{\ }$ ) over a vector to denote its contraction with the Dirac matrices, namely
\begin{equation}
\widehat{V}\equiv g_{\mu \nu}V^\mu \gamma^{\nu}, 
\end{equation}
with  $g_{\mu \nu}$=diag$\{1,-1,-1,-1\}$ and $\gamma^{\mu}$=$(\gamma^0,\gamma^1,\gamma^2,\gamma^3)$.
%It is obvious from Eq.~(\ref{ee1}) that there will be lots of matrix multiplications in the calculation which are quite time-consuming.

Since physical results are described by square of the amplitude, the conventional way to evaluate Eq.~(\ref{ee1}) is squaring it with its Hermitian conjugation, and summing over the fermion polarizations using Dirac equation. By doing so, the square of the amplitude can be expressed as
\begin{equation}
|{\cal M}|^2 \sim \sum\limits_{j}  \mathrm{Tr}S^{(n_{j_1})}_{j_1} \mathrm{Tr}S_{j_2}^{(n_{j_2})} \cdots \mathrm{Tr}S_{j_l}^{(n_{j_l})},
\end{equation}
where $S_i^{(n_{i})}$ denotes a string with $n_i$ Dirac matrices. Each of them is obtained from at least two fermion lines such that it is obvious there will be lots of matrix multiplications in the calculation which are quite time-consuming. Furthermore, in  multi-particle processes the number of total terms in Eq.~(\ref{ee1}) grows very rapidly, which leads to a tremendous number of terms after the squaring. Therefore, this conventional way is not applicable in multi-particle processes. 

An alternative way is to calculate the amplitude directly, before squaring it. A fermion line $F_k$ can be turned into a trace by moving the second spinor to the front, namely
\begin{equation}
F_k=\mathrm{Tr}(U_2\overline{U}_1 \widehat{V}_1\cdots \widehat{V}_m).
\end{equation}
With specific polarizations, $U_2\overline{U}_1$ is nothing but a $4\times4$ matrix and therefore can be expanded over $\gamma$-matrices. There are many different ways in the expansion which lead to different approaches (see e.g.~\cite{Vega:1995cc,Chang:1992bb,Bondarev:1997kf,Qiao:2003ue,Chen:2021tfj}).
%and the next step is to re-express 
%$u(n_2,k_2,s_2)  \otimes\overline{u}(n_1,k_1,s_1)$ by basic Dirac matrices. The re-expression can be many different ways and one can refer to these papers~\cite{Vega:1995cc,Chang:1992bb,Bondarev:1997kf,Qiao:2003ue,Chen:2021tfj}.

%Hence the fermion line are now decomposed to a combination of several traces:  
%\begin{equation}
%F_k\sim \sum_i \mathrm{Tr}S_i^{(n_i)}
%\label{ee2}
%\end{equation}
%There are many different ways to 

Another popular approach is the helicity amplitude method~\cite{Berends:1981rb,DeCausmaecker:1981jtq,Kleiss:1985yh,Gastmans:1990xh,Xu:1986xb,Gunion:1985vca,Hagiwara:1985yu}. For a certain helicity configuration of external particles, the amplitude can be 
simplified since many of the terms do not contribute. Meanwhile, amplitudes with different helicity configurations do not interfere with each other, hence the square of the total amplitude becomes straightforward.

%According to the number of Dirac matrices included, the strings on the r.h.s. of Eq.~(\ref{ee2}) can be separated into two groups, odd- and even-numbered. For a certain configuration of spinor helicities, only one of them will survive while the other will vanish thus Eq.~(\ref{ee2}) is further simplified.
%Meanwhile, the expansion of $U_2\overline{U}_1$ can also be reduced using helicity amplitude method.
% ({\color{red}more comments and citations}).

 In recent years, a new approach, based on on-shell recursion relations, becomes popular in the calculation of scattering amplitude. In this approach, the amplitude of multi-particle processes is constructed with blocks from on-shell amplitudes of fewer legs, which is much simpler than the original one. The most famous on-shell recursion relations are the ``BCFW recursion relations'' proposed by Britto, Cachazo, Feng and Witten in the calculation of gluon scattering~\cite{Britto:2004ap, Britto:2005fq}. The recursion relations are derived using complex deformation of the external momenta and calculating the residue of the deformed amplitude in the complex plane. 
Combined with the spinor helicity formalism, this approach has show great advantage in the scattering of massless particles. 
This approach cannot be directly applied to the processes with massive particles, since the momenta of massive particles cannot be written as a direct product of two spinors. There have been many efforts on this~\cite{Schwinn:2005pi, Schwinn:2006ca, Schwinn:2007ee, Craig:2011ws, Boels:2011zz, Arkani-Hamed:2017jhn, Aoude:2019tzn, Ballav:2020ese, Herderschee:2019dmc,Franken:2019wqr}.

In this paper, we introduce a new numerical algorithm for fermion lines calculation at tree-level. By avoiding most time-consuming matrix multiplications, this new approach could reduce the time used in the calculation of amplitude hence improve the efficiency of phase space integration and event generation, especially in the processes with many fermion lines.

%Although there are so many programs and methods, the calculation is still not easy. With the increase of the external particles, not only does the number of Feynman diagram grow up rapidly, but the result expression of the Dirac matrix string also expands, which lead to time consuming for the complete calculation.
%
%Taking the process $e^{+}e^{-}\rightarrow J/\psi c \overline{c}$ and FDC program as an example, the computation of the cross section is very fast, and one second is enough. However, if we add one gluon to the final particles, the time consumes becomes about 500 larger than before. Other programs have the same condition and one can see it in the following article. 

This paper is organized as follows: In Section~\ref{sec:method}, we detailed introduce our new algorithm, including the calculation of $\gamma$-matrix strings, the calculation of fermion lines and the calculation of the amplitudes with fermion lines. In Section~\ref{sec:compare}, the comparison with other programs is presented to show the advantage of the new algorithm which is implemented in our FDC package~\cite{Wang:2004du}.The summary is given in the last section.

\section{The approach}\label{sec:method}
\subsection{Notations and Conventions}
First, our approach is limited to tree level hence the dimension of space-time is always 4. 
And for the $\gamma$-matrices, the Dirac (standard) representation is used, namely
\begin{eqnarray}
\gamma^0 = \left( \begin{array}{ccc} 1 & 0 \\ 0 & -1 \end{array} \right),
\gamma^i = \left( \begin{array}{ccc} 0 & \sigma^i \\ -\sigma^i & 0 
\end{array} \right),
\gamma^5 = \left( \begin{array}{ccc} 0 & 1 \\ 1 & 0 \end{array} \right),\quad
\label{p0}
\end{eqnarray}
with $i=1,2,3$.  $\sigma_i$ are the Pauli matrices:
\begin{eqnarray}
\sigma^1 = \left( \begin{array}{ccc} 0 & 1 \\ 1 & 0 \end{array} \right),
\sigma^2 = \left( \begin{array}{ccc} 0 & -i \\ i & 0 \end{array} \right), 
\sigma^3 =  \left( \begin{array}{ccc} 1 & 0 \\ 0 & -1 \end{array} \right). \quad
\end{eqnarray}
And the contraction of a 4-dimensional vector $p^\mu=(p^0,\vec{p})$ with the $\gamma$-matrices takes the form of
%\begin{eqnarray}
\begin{equation}
\widehat{p}=p^0\gamma^0-\vec{p}\cdot \vec{\gamma}=
\left( \begin{array}{ccc}
p^0 & -\vec{p}\cdot\vec{\sigma}  \\ 
\vec{p}\cdot\vec{\sigma} & -p^0
\end{array} \right).
\label{eqn:pslash}
\end{equation}
%\end{eqnarray}
The string with $n$ $\gamma$-matrices is defined as
\begin{equation}
S^{(n)}(p_n,p_{n-1},\cdots,p_1)=\widehat{p}_n\widehat{p}_{n-1}\cdots\widehat{p}_1.
\label{eqn:sn0}
\end{equation}
The order of superscripts in the vectors has been reversed for convenience. 
%And we use $S^{(2n)}$ and $S^{(2n+1)}$ to denotes even- and odd-numbered strings ($n\ge0$), respectively.
%
%In the following article, for convenience, we define 
%\begin{eqnarray}
%B_m \equiv \widehat{p}_m\dotsb\widehat{p}_2\widehat{p}_1,
%\end{eqnarray}
%where m can be any non-negative integer.
%\subsection{Simplification of spinor}

The Dirac spinors for fermion and anti-fermion are defined as
\begin{eqnarray}
u(k,s)&\equiv& U(+,k,s)=\dfrac{\widehat{k}+m}{\sqrt{k^0+m}}u_0 (+,s),\nonumber \\
v(k,s)&\equiv& U(-,k,s)=\dfrac{-\widehat{k}+m}{\sqrt{k^0+m}}u_0 (-,s),
\label{eqn:uv}
\end{eqnarray}
with
\begin{equation}
u_0(+,s)\equiv\left(\begin{array}{ccc} \chi_s \\  0 \end{array}\right), \quad
u_0(-,s)\equiv \left(\begin{array}{ccc} 0 \\ \chi_s \end{array}\right),  
\label{p12}
\end{equation}
and
\begin{equation}
\chi_1=\left(\begin{array}{ccc} 1 \\  0 \end{array}\right), \quad
\chi_2=\left(\begin{array}{ccc} 0 \\ 1 \end{array}\right).  
\end{equation}

%%\begin{eqnarray}
%\begin{equation}
%\chi^s=\begin{cases}
%\left( \begin{array}{ccc} 1 & 0 \end{array} \right)^T,  &s=1\\
%\left( \begin{array}{ccc} 0 & 1 \end{array} \right)^T, &s=2
%\end{cases}
%\end{equation}
%\end{eqnarray}
Here the symbols $\pm$ in $U$ and $u_0$ are used to distinguish fermion and anti-fermion, and $s$ is the polarization of the particle.

%\begin{eqnarray}
%\chi_s=\begin{cases}
%\left( \begin{array}{ccc} 1 & 0 \end{array} \right)^T,  &s=1\\
%\left( \begin{array}{ccc} 0 & 1 \end{array} \right)^T, &s=2
%\end{cases}
%\end{eqnarray}
%where $n=+$ or $-$ represents particle or antiparticle, and $s=1$ or 2 represents up or down polarization.
%
%Substituting the new form (\ref{p12}) in Eq. (\ref{p8}), one can get 
%\begin{eqnarray}
%\begin{split}
%M
%&=\mu(n_1,s_1)^{+} \widehat{k}_1^{\prime+}\gamma_0 B_m \widehat{k}_2^{\prime}\mu(n_2,s_2) \\
%&=\mu(n_1,s_1)^{+}\gamma_0\widehat{k}_1^{\prime} B_m \widehat{k}_2^{\prime}\mu(n_2,s_2) \\
%&=\mu(n_1,s_1)^{+}\gamma_0 A_n\mu(n_2,s_2)
%\end{split}
%\label{p10}
%\end{eqnarray}
%where we have defined 
%\begin{eqnarray}
%A_n \equiv \widehat{k}_1^{\prime} B_m \widehat{k}_2^{\prime},\quad n=m+2.
%\end{eqnarray}
%In general, one can redefine
%\begin{eqnarray}
%A_n \equiv \widehat{p}_n\dotsb\widehat{p}_2\widehat{p}_1,\quad n=2,3,\dotsb.
%\end{eqnarray} 
%Meanwhile, if there is a $\gamma^5$ in front of $B_m$
%\begin{eqnarray}
%M=\overline{u}(n_1,k_1,s_1) \gamma^5 B_m u(n_2,k_2,s_2), 
%\end{eqnarray}
%the final result equals
%\begin{eqnarray}
%M=-\mu(n_1,s_1)^{+}\gamma_0 \gamma^5 A_n\mu(n_2,s_2).
%\label{p16}
%\end{eqnarray}

\subsection{Calculation of $S^{(n)}$}
\label{sec:sn}
In this subsection, we introduce how we calculate $S^{(n)}$, the string with $n$ $\gamma$-matrices  recursively. In the calculation of the amplitude there might be $\gamma^5$ among other $\gamma$-matrices. Due to its anti-commutativity with all the $\gamma$-matrices, all the $\gamma^5$ can be moved outside of $S^{(n)}$ by adding a possible factor $-1$. Meanwhile we suppose all the Lorentz indices have been summed over in the string therefore all the $\gamma$-matrices in $S^{(n)}$ are now contracted with certain vectors, i.e. it takes exactly the form of Eq.~(\ref{eqn:sn0}).
%Here we mainly simplify the situation of more than one $\widehat{p}_i$ and finally we can find all these kinds of formula can be expressed as a recursive form related to odevity of its number.

\subsubsection{$n=2$}
We will show later that the cases of $n=0, 1$ are not needed. Here we just skip them and start from $n=2$. According to Eq.~(\ref{eqn:pslash}), $S^{(2)}$ can be rewritten as
\begin{eqnarray}
\begin{split}
&S^{(2)}(p_2,p_1) \\
=& \left( \begin{array}{rrr}
p_2^0 & -\vec{p}_2\cdot\vec{\sigma}  \\ 
\vec{p}_2\cdot\vec{\sigma} & -p_2^0
\end{array} \right)
\left( \begin{array}{rrr}
p_1^0 & -\vec{p}_1\cdot\vec{\sigma}  \\ 
\vec{p}_1\cdot\vec{\sigma} & -p_1^0
\end{array} \right) \\
%---------------------
=&\left( \begin{array}{rrr}
p_2^0 p_1^0-(\vec{p}_2\cdot\vec{\sigma})( \vec{p}_1\cdot\vec{\sigma})& 
p_1^0 (\vec{p}_2\cdot\vec{\sigma})-p_2^0(\vec{p}_1\cdot\vec{\sigma})  \\[2mm]
p_1^0 (\vec{p}_2\cdot\vec{\sigma})-p_2^0(\vec{p}_1\cdot\vec{\sigma})& 
p_2^0 p_1^0-(\vec{p}_2\cdot\vec{\sigma} )(\vec{p}_1\cdot\vec{\sigma})
\end{array} \right).
\end{split}
\end{eqnarray}
%Here we calculate $A_2=\widehat{p}_2\widehat{p}_1$ and $p_2$, $p_1$ can represent all $p_i$ conditions. Using specific form of $\widehat{p}$, one can get
%\begin{eqnarray}
%\begin{split}
%A_2&=
%\left( \begin{array}{ccc}
%p_2^0 & -\vec{p_2}\cdot\vec{\sigma}  \\ 
%\vec{p_2}\cdot\vec{\sigma} & -p_2^0
%\end{array} \right)
%\left( \begin{array}{ccc}
%p_1^0 & -\vec{p_1}\cdot\vec{\sigma}  \\ 
%\vec{p_1}\cdot\vec{\sigma} & -p_1^0
%\end{array} \right) \\
%&=\left( \begin{array}{ccc}
%p_2^0 p_1^0-\vec{p_2}\cdot\vec{\sigma} 
%\vec{p_1}\cdot\vec{\sigma}
%& 
%p_1^0 \vec{p_2}\cdot\vec{\sigma}
%-p_2^0\vec{p_1}\cdot\vec{\sigma}  
%\\
%p_1^0 \vec{p_2}\cdot\vec{\sigma}
%-p_2^0\vec{p_1}\cdot\vec{\sigma}  
%& 
%p_2^0 p_1^0-\vec{p_2}\cdot\vec{\sigma} \vec{p_1}\cdot\vec{\sigma}
%\end{array} \right).
%\end{split}
%\end{eqnarray}
Using the identity of Pauli matrices 
\begin{equation}
\sigma^i\sigma^j=i\varepsilon^{ijk}\sigma^k+\delta^{ij},
\end{equation}
it is easy to find
\begin{equation}
\begin{split}
(\vec{p}_2\cdot\vec{\sigma})(\vec{p}_1\cdot\vec{\sigma})&=p_2^i\sigma^ip_1^j\sigma^j\\
&=p_2^ip_1^j(i\varepsilon^{ijk}\sigma^k+\delta^{ij})\\
%\sigma^i\sigma^j
&=i\varepsilon^{ijk}p_2^ip_1^j\sigma^k+\vec{p}_2\cdot\vec{p}_1.
\end{split}
\end{equation}
%By the calculation of the following formula 
%\begin{eqnarray}
%\vec{p_2}\cdot\vec{\sigma} 
%\vec{p_1}\cdot\vec{\sigma}=p_2^i\sigma^ip_1^j\sigma^j
%=p_2^ip_1^j\sigma^i\sigma^j
%=i\varepsilon_{ijk}p_2^ip_1^j\sigma^k+\vec{p_2}\cdot\vec{p_1},
%\end{eqnarray}
%where the following properties of Pauli matrices are used:
%$$
%\sigma_i\sigma_j=i\varepsilon_{ijk}\sigma_k+\delta_{ij},
%$$
Hence
\begin{eqnarray}
\begin{split}
&S^{(2)}(p_2,p_1)\\
=&\left( \begin{array}{rrr}
p_2\cdot p_1-i\varepsilon^{ijk}p_2^ip_1^j\sigma^k& 
(p_1^0 \vec{p}_2-p_2^0\vec{p}_1)\cdot\vec{\sigma}  \\[2mm]
(p_1^0 \vec{p}_2-p_2^0\vec{p}_1)\cdot\vec{\sigma} & 
p_2\cdot p_1-i\varepsilon^{ijk}p_2^ip_1^j\sigma^k
\end{array} \right).
\end{split}
\label{p1} 
\end{eqnarray}
%
%one can continue the calculation and obtain 
%\begin{eqnarray}
%A_2=
%\left( \begin{array}{ccc}
%p_2\cdot p_1-i\varepsilon_{ijk}p_2^ip_1^j\sigma^k
%& 
%(p_1^0 \vec{p_2}-p_2^0\vec{p_1})\cdot\vec{\sigma}  
%\\
%(p_1^0 \vec{p_2}-p_2^0\vec{p_1})\cdot\vec{\sigma} 
%& 
%p_2\cdot p_1-i\varepsilon_{ijk}p_2^ip_1^j\sigma^k
%\end{array} \right).
%\label{p1} 
%\end{eqnarray}

On the other hand, by introducing two new vectors $p_a$ and $p_b$ as
\begin{equation}
\begin{split}
p_a^{(2),\mu}(p_2,p_1)&\equiv(p_2\cdot p_1, -p_2^0\vec{p}_1+p_1^0\vec{p}_2),\\
p_b^{(2),\mu}(p_2,p_1)&\equiv(0, \varepsilon^{ijk}p_2^ip_1^j),
\end{split}
\label{p13}
\end{equation}
 $S^{(2)}$ in Eq.~(\ref{p1}) can be further expressed as
\begin{equation}
S^{(2)}(p_2,p_1)=\widehat{p}_{a,\{2,1\}}^{(2)}\gamma^0-i\gamma^5\widehat{p}_{b,\{2,1\}}^{(2)}\gamma^0.
\label{p3}
\end{equation}
Here the superscript $(2)$ in $p_a$ and $p_b$ denotes the vector contains two arguments, which is same as $S^{(n)}$. 
And we have used the abbreviation to compact the expression
\begin{equation}
{p}_{a(b),\{i_n,i_{n-1},\cdots,i_1\}}^{(n)}\equiv{p}_{a(b)}^{(n)}(p_{i_n},p_{i_{n-1}},\cdots,p_{i_1}).
\end{equation}

%Correspondingly,
%In Eq.~(\ref{p3}), $p_a$ and $p_b$ on the r.h.s. have same arguments as $S^{(2)}$ hence are ignored in the equation.

%On the other hand, one can define two new $p$ donated by $p_a$, $p_b$, which are
%\begin{eqnarray}
%\begin{split}
%&p_a=(p_2\cdot p_1, -p_2^0\vec{p_1}+p_1^0\vec{p_2}),\\
%&p_b=(0, \quad \varepsilon_{ijk}p_2^ip_1^j).
%\label{p13}
%\end{split}
%\end{eqnarray}
%Using Eq. (\ref{p13}), one can rewrite the form of Eq. (\ref{p1}), which equals
%\begin{eqnarray}
%\begin{split}
%A_2=\widehat{p}_a\gamma^0-i\gamma^5\widehat{p}_b\gamma^0,
%\label{p3}
%\end{split}
%\end{eqnarray} 
%and that is the formula we need.
%Meanwhile, if there is a $\gamma^5$ in front of $A_2$, one can get
%\begin{eqnarray}
%\gamma^5 A_2=-i(\widehat{p}_b\gamma^0+i\gamma^5\widehat{p}_a\gamma^0).
%\end{eqnarray}

\subsubsection{$n=3$}
The calculation of $S^{(3)}$ is straightforward using Eq.~(\ref{p3})
\begin{eqnarray}
\begin{split}
&S^{(3)}(p_3,p_2,p_1) \\
=&\widehat{p}_3S^{(2)}(p_2,p_1) \\
=&\widehat{p}_3\left(\widehat{p}_{a,\{2,1\}}^{(2)}\gamma^0-i\gamma^5\widehat{p}_{b,\{2,1\}}^{(2)}\gamma^0\right) \\
=&S^{(2)}\left(p_3,{p}_{a,\{2,1\}}^{(2)}\right)\gamma^0+i\gamma^5S^{(2)}\left(p_3,{p}_{b,\{2,1\}}^{(2)}\right)\gamma^0.
%\widehat{p}_3\widehat{p}_a+i\gamma^5\widehat{p}_3\widehat{p}_b\gamma^0.
\label{p4}
\end{split}
\end{eqnarray}
The two $S^{(2)}$ on the r.h.s. of Eq.~(\ref{p4}) can be obtained by using Eq.~(\ref{p3}) again:
\begin{equation}
\begin{split}
S^{(2)}\left(p_3,{p}_{a(b),\{2,1\}}^{(2)}\right)
=&\widehat{p}^{(2)}_{a}\left(p_3,{p}_{a(b),\{2,1\}}^{(2)}\right)\gamma^0 \\
&-i\gamma^5\widehat{p}^{(2)}_{b}\left(p_3,{p}_{a(b),\{2,1\}}^{(2)}\right)\gamma^0.
\end{split}
\end{equation}
%\begin{eqnarray}
%\begin{split}
%&S^{(2)}\left(p_3,{p}_{a(b),\{2,1\}}^{(2)}\right)=\widehat{p}^{(2)}_{a}\left(p_3,{p}_{a(b),\{2,1\}}^{(2)}\right)\gamma^0
%-i\gamma^5\widehat{p}^{(2)}_{b}\left(p_3,{p}_{a(b),\{2,1\}}^{(2)}\right)\gamma^0 ,\\
%&S^{(2)}\left(p_3,{p}_{b,\{2,1\}}^{(2)}\right)=\widehat{p}^{(2)}_{a}\left(p_3,{p}_{b,\{2,1\}}^{(2)}\right)\gamma^0
%-i\gamma^5\widehat{p}^{(2)}_{b}\left(p_3,{p}_{b,\{2,1\}}^{(2)}\right)\gamma^0.	
%\end{split}
%\end{eqnarray}
Inserting them into Eq.~(\ref{p4}) one finds
%\begin{equation}
\begin{eqnarray}
\begin{split}
&S^{(3)}(p_3,p_2,p_1) \\
=&\left[\widehat{p}^{(2)}_{a}\left(p_3,{p}_{a,\{2,1\}}^{(2)}\right)+\widehat{p}^{(2)}_{b}\left(p_3,{p}_{b,\{2,1\}}^{(2)}\right)\right] \\
&
-i\gamma^5\left[\widehat{p}^{(2)}_{b}\left(p_3,{p}_{a,\{2,1\}}^{(2)}\right)-\widehat{p}^{(2)}_{a}\left(p_3,{p}_{b,\{2,1\}}^{(2)}\right)\right].
\end{split}
\end{eqnarray}
If introducing two new vectors as
%\begin{equation}
\begin{eqnarray}
p^{(3)}_a(p_3,p_2,p_1)&\equiv&{p}^{(2)}_{a}\left(p_3,{p}_{a,\{2,1\}}^{(2)}\right)+{p}^{(2)}_{b}\left(p_3,{p}_{b,\{2,1\}}^{(2)}\right),
\nonumber \\
p^{(3)}_b(p_3,p_2,p_1)&\equiv&{p}^{(2)}_{b}\left(p_3,{p}_{a,\{2,1\}}^{(2)}\right)-{p}^{(2)}_{a}\left(p_3,{p}_{b,\{2,1\}}^{(2)}\right),
\nonumber \\
\label{eqn:pab3}
\end{eqnarray}
$S^{(3)}$ can be expressed as
\begin{equation}
S^{(3)}(p_3,p_2,p_1)=\widehat{p}^{(3)}_{a,\{3,2,1\}}-i\gamma^5\widehat{p}^{(3)}_{b,\{3,2,1\}}.
\label{eqn:s3}
\end{equation}

\subsubsection{$n=4$ and general formula of $S^{(n)}$}
The calculation for the case of $n=4$ is similar as the case of $n=3$. 
Using Eq.~(\ref{eqn:s3}), $S^{(4)}$ is expressed by two other $S^{(2)}$ as
\begin{eqnarray}
\begin{split}
&S^{(4)}(p_4,p_3,p_2,p_1) \\
=&\widehat{p_4}\left[\widehat{p}^{(3)}_{a,\{3,2,1\}}-i\gamma^5\widehat{p}^{(3)}_{b,\{3,2,1\}}\right]  \\
=&S^{(2)}\left(p_4,{p}^{(3)}_{a,\{3,2,1\}}\right)+i\gamma^5S^{(2)}\left(p_4,{p}^{(3)}_{b,\{3,2,1\}}\right) .
\end{split}
\label{eqn:s4_1}
\end{eqnarray}
These $S^{(2)}$ are then obtained by using Eq.~(\ref{p3}) as
\begin{equation}
\begin{split}
S^{(2)}\left(p_4,{p}_{a(b),\{3,2,1\}}^{(3)}\right)
=&\widehat{p}^{(2)}_{a}\left(p_4,{p}_{a(b),\{3,2,1\}}^{(3)}\right)\gamma^0 \\
&-i\gamma^5\widehat{p}^{(2)}_{b}\left(p_4,{p}_{a(b),\{3,2,1\}}^{(3)}\right)\gamma^0 .
\end{split}
\end{equation}
%\begin{eqnarray}
%\begin{split}
%&S^{(2)}\left(p_4,{p}_{a(b),\{3,2,1\}}^{(3)}\right)=\widehat{p}^{(2)}_{a}\left(p_4,{p}_{a(b),\{3,2,1\}}^{(3)}\right)\gamma^0
%-i\gamma^5\widehat{p}^{(2)}_{b}\left(p_4,{p}_{a(b),\{3,2,1\}}^{(3)}\right)\gamma^0 ,\\
%&S^{(2)}\left(p_4,{p}_{b,\{3,2,1\}}^{(3)}\right)=\widehat{p}^{(2)}_{a}\left(p_4,{p}_{b,\{3,2,1\}}^{(3)}\right)\gamma^0
%-i\gamma^5\widehat{p}^{(2)}_{b}\left(p_4,{p}_{b,\{3,2,1\}}^{(3)}\right)\gamma^0.	
%\end{split}
%\end{eqnarray}
Inserting them into Eq.~(\ref{eqn:s4_1}) gives 
\begin{eqnarray}
\begin{split}
&S^{(4)}(p_4,p_3,p_2,p_1)\\
=&\left[\widehat{p}^{(2)}_{a}\left(p_4,{p}_{a,\{3,2,1\}}^{(3)}\right)+\widehat{p}^{(2)}_{b}\left(p_4,{p}_{b,\{3,2,1\}}^{(3)}\right)\right]\gamma^0 \\
&-i\gamma^5\left[\widehat{p}^{(2)}_{b}\left(p_4,{p}_{a,\{3,2,1\}}^{(3)}\right)-\widehat{p}^{(2)}_{a}\left(p_4,{p}_{b,\{3,2,1\}}^{(3)}\right)\right]\gamma^0 \\
=&\widehat{p}_{a,\{4,3,2,1\}}^{(4)}\gamma^0-i\gamma^5\widehat{p}_{b,\{4,3,2,1\}}^{(4)}\gamma^0,
\end{split}
\label{eqn:s4}
\end{eqnarray}
where the two new vectors are defined as
\begin{eqnarray}
p^{(4)}_a(p_4,p_3,p_2,p_1)&\equiv&{p}^{(2)}_{a}\left(p_4,{p}_{a,\{3,2,1\}}^{(3)}\right)+{p}^{(2)}_{b}\left(p_4,{p}_{b,\{3,2,1\}}^{(3)}\right),
\nonumber \\
p^{(4)}_b(p_4,p_3,p_2,p_1)&\equiv&{p}^{(2)}_{b}\left(p_4,{p}_{a,\{3,2,1\}}^{(3)}\right)-{p}^{(2)}_{a}\left(p_4,{p}_{b,\{3,2,1\}}^{(3)}\right).
\nonumber \\
\label{eqn:pab4}
\end{eqnarray}
It can be seen that $S^{(4)}$ in Eq.~(\ref{eqn:s4}) has the same form as $S^{(2)}$, therefore $S^{(5)}$ will have the same form as $S^{(3)}$.
Generalizing this to even larger $n$, we come to the general formula of $S^{(n)}$ ($n\ge2$):
\begin{widetext}
\begin{eqnarray}
S^{(n)}(p_n,p_{n-1},\cdots,p_1)=
\begin{cases}
\widehat{p}_{a,\{n,n-1,\cdots,1\}}^{(n)}\gamma^0-i\gamma^5\widehat{p}_{b,\{n,n-1,\cdots,1\}}^{(n)}\gamma^0 &\qquad \text{for even $n$}\\[2mm]
\widehat{p}_{a,\{n,n-1,\cdots,1\}}^{(n)}-i\gamma^5\widehat{p}_{b,\{n,n-1,\cdots,1\}}^{(n)}  &\qquad \text{for odd $n$}
\end{cases}
\label{p17}
\end{eqnarray} 
\end{widetext}
which can be further rewritten as
\begin{equation}
\begin{split}
&S^{(n)}(p_n,p_{n-1},\cdots,p_1) \\
=&\left(
\widehat{p}_{a,\{n,n-1,\cdots,1\}}^{(n)}-i\gamma^5\widehat{p}_{b,\{n,n-1,\cdots,1\}}^{(n)}\right)(\gamma^0)^{n+1}.
\end{split}
\label{eqn:sn}
\end{equation}
Meanwhile, from Eqs.~(\ref{eqn:pab3}) and (\ref{eqn:pab4}) it can seen that $p_a^{(n)}$ and $p_b^{(n)}$ ($n\ge3$) always have the same form no matter $n$ is odd or even. It can be obtained recursively as
\begin{widetext}
\begin{eqnarray}
\begin{split}
p^{(n)}_a(p_n,p_{n-1},\cdots,p_1)
\equiv&{p}^{(2)}_{a}\left(p_n,{p}_{a,\{n-1,\cdots,1\}}^{(n-1)}\right)+{p}^{(2)}_{b}\left(p_n,{p}_{b,\{n-1,\cdots,1\}}^{(n-1)}\right),\\
p^{(n)}_b(p_n,p_{n-1},\cdots,p_1)
\equiv&{p}^{(2)}_{b}\left(p_n,{p}_{a,\{n-1,\cdots,1\}}^{(n-1)}\right)-{p}^{(2)}_{a}\left(p_n,{p}_{b,\{n-1,\cdots,1\}}^{(n-1)}\right),
\end{split}
\label{eqn:pabn}
\end{eqnarray}
\end{widetext}
while $p_{a(b)}^{(2)}$ is defined in Eq.~(\ref{p13}).

Let us further investigate the structure of $S^{(n)}$. 
Eq.~(\ref{eqn:sn}) can be taken as another definition of $S^{(n)}$, in which the arguments are no longer $p_i$ but $p_a$ and $p_b$. For a fixed queue of $p_i$, $p_a$ and $p_b$ can be obtained with Eq.~(\ref{eqn:pabn}) and then the corresponding $S^{(n)}$ is expressed as
\begin{equation}
S^{(n)}(p_a,p_b)=\left(
\widehat{p}_{a}-i\gamma^5\widehat{p}_{b}\right)(\gamma^0)^{n+1}\equiv \widetilde{S}^{(n)}(p_a,p_b)(\gamma^0)^{n+1}.
\label{eqn:sntilde}
\end{equation}
A tilde is used over $S^{(n)}$ to denotes the part without $\gamma^0$.
With Eqs.~(\ref{p0}) and (\ref{eqn:pslash}), it is easy to find that $\widetilde{S}^{(n)}$ can be expressed by four blocks as
\begin{equation}
\widetilde{S}^{(n)}(p_a,p_b)=
\left( \begin{array}{ccc}
A^{(++)} & A^{(+-)} \\ 
A^{(-+)} & A^{(--)}
\end{array} \right), 
\label{eqn:snblock}
\end{equation}
with 
\begin{eqnarray}
A^{(++)}&=-A^{(--)}=p_a^0-i\vec{p}_b\cdot \vec{\sigma} ,\nonumber \\
A^{(+-)}&=-A^{(-+)}=ip_b^0-\vec{p}_a\cdot \vec{\sigma} .
\end{eqnarray}
Only two of them are independent.

Also, from Eq.~(\ref{eqn:sntilde}) it is straightforward to derive the product of $\gamma^5$ and $S^{(n)}$:
\begin{equation}
\begin{split}
\gamma^5S^{(n)}(p_a,p_b)&=
\left(\gamma^5\widehat{p}_{a}-i\widehat{p}_{b}\right)(\gamma^0)^{n+1} \\
&=(-i)\left(\widehat{p}_{b}+i\gamma^5\widehat{p}_{a}\right)(\gamma^0)^{n+1}\\
&=-iS^{(n)}(p_b,-p_a).
\end{split}
%\equiv \widetilde{S}^{(n)}(p_a,p_b)(\gamma^0)^{n+1}
\label{eqn:g5sn}
\end{equation}

\subsection{Calculation of fermion lines}
In this subsection, we introduce how we calculate fermion lines with the string $S^{(n)}$ we obtained above. A fermion line with two external fermions usually can be written as
\begin{equation}
\begin{split}
&F(z_1, k_1, s_1;z_2, k_2, s_2; n)\\
=&\overline{U}(z_1,k_1,s_1)S^{(n)}(p_n,p_{n-1},\cdots,p_1)U(z_2,k_2,s_2).
\end{split}
\label{eqn:f0}
\end{equation}
In this expression ${U}(z,k,s)$, whose definition can be found in Eq.~(\ref{eqn:uv}),  is used as the spinor of fermion and anti-fermion at the same time. Here $z$ can be $+$ or $-$ which stands for fermion or anti-fermion, respectively.$k$ and $s$ are the momentum and polarization of the particle. It should be pointed out that for a specific process, all the $z_i$ are fixed according to external particles. $n$ in the arguments of $F$ denotes a string of $n$ $\gamma$-matrices, but we have removed the detailed list of vectors for convenience. 

In the calculation of amplitude there might be $\gamma^5$ inside a fermion line. However, due to its anti-commutativity with all four $\gamma$-matrices, it can always be moved in front of $S^{(n)}$. Furthermore according to Eq.~(\ref{eqn:g5sn}), $\gamma^5S^{(n)}$ can be obtained by a substitution in the arguments of $S^{(n)}$. Hence there is no particular need to discuss how to calculate fermion lines with $\gamma^5$. 

\subsubsection{a trick in the spinors}
First we introduce a trick in the spinors. As shown in Eq.~(\ref{eqn:uv}), the relationship between $U(z, k, s)$ and $u_0(z,s)$ can be expressed as
\begin{equation}
U(z,k,s)=\dfrac{z\widehat{k}+m}{\sqrt{k^0+m}}u_0 (z,s).
\end{equation}
If we introduce a new vector which is defined as
\begin{equation}
k^{\prime\mu}\equiv\dfrac{1}{\sqrt{k^0+m}}({k^0+m},\vec{k}),
\end{equation}
it is easy to prove 
\begin{equation}
U(z,k,s)=z\widehat{k}^\prime u_0 (z,s), \quad \overline{U}(z,k,s)=z \overline{u}_0 (z,s)\widehat{k}^\prime.
\label{eqn:ukprime}
\end{equation}

This trick is very useful in case of massive fermions, since it prevents the term $z\widehat{k}+m$ being separated. In a process with $n$ massive fermions, this trick can reduce the number of total terms by a factor of $1/2^n$. 

Meanwhile it can be observed that in the Dirac presentation, $u_0$ defined in Eq.~(\ref{p12}) is an eigenstates of $\gamma^0$ with the eigenvalue $z$, namely 
\begin{equation}
\gamma^0 u_0(z,s)=z u_0(z, s), \quad u_0(z,s)^\dagger \gamma^0 =z u_0(z, s)^\dagger.
\label{eqn:u0g0}
\end{equation}
%
%
%
%As introduced above, one term for a fermion line in Feynman amplitude can be generally expressed as 
%\begin{eqnarray}
%\begin{split}
%M
%&=\overline{u}(n_1,k_1,s_1) B_m u(n_2,k_2,s_2)  \\
%&=u^{+}(n_1,k_1,s_1)\gamma^0 B_m u(n_2,k_2,s_2),
%\end{split} 
%\label{p8}
%\end{eqnarray}
%where 
%\begin{eqnarray}
%u(n,k,s)=
%\begin{cases}
%\dfrac{\widehat{k}+m}{\sqrt{k^0+m}}\mu (+,s)=\widehat{k}^{\prime}\mu(+,s), &n=+ \\
%\dfrac{\widehat{k}-m}{\sqrt{k^0+m}}\mu (-,s)=\widehat{k}^{\prime}\mu(-,s), &n=- 
%\end{cases}
%\label{p9}
%\end{eqnarray}
%and
%\begin{eqnarray}
%\mu(+,s)=\left(\begin{array}{ccc} \chi_s \\  0 \end{array}\right), & \quad
%\mu(-,s)=\left(\begin{array}{ccc} 0 \\ \chi_s \end{array}\right),  & \quad
%k^{\prime\mu}=(\sqrt{k^0+m},\quad \dfrac{\vec{k}}{\sqrt{k^0+m}}),
%\label{p12}
%\end{eqnarray}

\subsubsection{calculation of fermion lines without Lorentz indices}
Now we turn back to the calculation of fermions line.
Here we still suppose that all the Lorentz indices have been summed over in the string of $\gamma$-matrices hence no more Lorentz indices are remained.

Using Eqs.~(\ref{eqn:ukprime}) and (\ref{eqn:u0g0}), it is easy to find
% into Eq.~(\ref{eqn:f0}), 
\begin{eqnarray}
&&F(z_1, k_1, s_1;z_2, k_2, s_2;n) \nonumber \\
&=&\overline{U}(z_1,k_1,s_1)S^{(n)}(p_n,p_{n-1},\cdots,p_1)U(z_2,k_2,s_2)  \\
&=&z_1z_2\overline{u}_0(z_1,s_1)\widehat{k}_1^\prime S^{(n)}(p_n,p_{n-1},\cdots,p_1) \widehat{k}_2^\prime u_0(z_2,s_2) \nonumber \\
&=&z_1z_2\overline{u}_0(z_1,s_1)S^{(n+2)}(k_1^\prime, p_n,p_{n-1},\cdots,p_1, k_2^\prime)u_0(z_2,s_2). \nonumber
\label{eqn:f1}
\end{eqnarray}
The momenta are separated from the spinors and merged to the string of $\gamma$-matrices $S^{(n+2)}$. This procedure can be done to all the fermion lines, which indicates all the $S^{(n)}$ we need to calculate contain at least two $\gamma$-matrices. This is the reason why in Sec.~\ref{sec:sn} only the cases of $n\ge2$ are considered. 

The r.h.s. of Eq.~(\ref{eqn:f1}) can further be evaluated with Eqs.~(\ref{eqn:sntilde}) and (\ref{eqn:snblock}) as
\begin{eqnarray}
&&F(z_1, k_1, s_1;z_2, k_2, s_2;n) \nonumber \\
&=&z_1z_2{u}_0(z_1,s_1)^\dagger\gamma^0\widetilde{S}^{(n+2)}(\gamma^0)^{n+3}u_0(z_2,s_2) \nonumber \\[1mm]
&=&z_1^2z_2^{n+4}{u}_0(z_1,s_1)^\dagger
\left( \begin{array}{ccc}A^{(++)} & A^{(+-)} \\ A^{(-+)} & A^{(--)}\end{array} \right)u_0(z_2,s_2) \nonumber \\ %[2mm]
&=&z_2^{n}\chi_{s_1}^\dagger A^{(z_1z_2)}\chi_{s_2} \nonumber  \\[2mm]
&=&z_2^n A^{(z_1z_2)}_{s_1s_2},
\label{eqn:f2}
\end{eqnarray}
where Eq.~(\ref{eqn:u0g0}) and $z_1^2=z_2^2=1$ have been used. 

As aforementioned, $z_{1,2}$ is fixed for a specific process, hence only one of the four blocks in $\widetilde{S}^{(n+2)}$ is needed. 
Since $\widetilde{S}^{(n+2)}$ is totally decided by $p_a$ and $p_b$, the calculation of the fermion line is now converted into the calculation of $p_a$ and $p_b$, which can be recursively without any matrix multiplications.

Furthermore, four elements of the block $A^{(z_1z_2)}$ corresponds to the four different configurations of $\{s_1,s_2\}$, respectively.  This means that the results for all possible polarization configurations can be obtained via a single calculation, which can improve the efficiency of calculation greatly.
\\

\subsubsection{calculation of fermion lines with one Lorentz index}
\label{sec:Fu}
So far we have supposed that all the Lorentz indices have been summed over in fermion lines, but that is not enough for the calculation of amplitude. In this section, we introduce how to calculate fermion lines with one Lorentz index, which take the form of 
\begin{equation}
\begin{split}
&F^\mu(z_1, k_1, s_1;z_2, k_2, s_2; n_1,n_2) \\
=&\overline{U}(z_1,k_1,s_1)S^{(n_1)}(p_{n_1},\cdots,p_1)\gamma^\mu \\ &S^{(n_2)}(q_{n_2},\cdots,q_1)U(z_2,k_2,s_2).
\end{split}
\label{eqn:fu0}
\end{equation}
It is a Lorentz vector with four components. 
In order to calculate these four components, four auxiliary vectors are introduced as follows
\begin{eqnarray}
r^\mu_0=\{1 , 0  ,  0  , 0\}, \quad r^\mu_1=\{0 , -1 , 0 , 0\}, \nonumber \\ 
r^\mu_2=\{0 , 0 , -1 , 0\},  \quad r^\mu_3=\{0 , 0 , 0 , -1\}.
\end{eqnarray}
It is obvious that 
\begin{equation}
\gamma^0=\widehat{r}_0, \quad \gamma^i=\widehat{r}_i .
\end{equation}
Hence
\begin{widetext}
\begin{eqnarray}
F^0(z_1, k_1, s_1;z_2, k_2, s_2; n_1,n_2) 
&=&\overline{U}(z_1,k_1,s_1)S^{(n_1)}(p_{n_1},\cdots,p_1)\gamma^0S^{(n_2)}(q_{n_2},\cdots,q_1)U(z_2,k_2,s_2)\nonumber \\
&=&\overline{U}(z_1,k_1,s_1)S^{(n_1)}(p_{n_1},\cdots,p_1)\widehat{r}_0S^{(n_2)}(q_{n_2},\cdots,q_1)U(z_2,k_2,s_2)\nonumber \\
%&=&\overline{U}(z_1,k_1,s_1)S^{(n_1)}\{\widehat{r}_0,\widehat{r}_1,\widehat{r}_2,\widehat{r}_3\} S^{(n_2)}U(z_2,k_2,s_2) \\
&=&\overline{U}(z_1,k_1,s_1)S^{(n_1+n_2+1)}(p_{n_1},\cdots,p_1,{r}_0,q_{n_2},\cdots,q_1)U(z_2,k_2,s_2).
\nonumber
\label{eqn:fu1}
\end{eqnarray}
\end{widetext}
It is exactly the result of a fermion line without any Lorentz indices. Similarly, the other three components are obtained as
\begin{widetext}
\begin{equation}
F^i(z_1, k_1, s_1;z_2, k_2, s_2; n_1,n_2)
=\overline{U}(z_1,k_1,s_1)S^{(n_1+n_2+1)}(p_{n_1},\cdots,p_1,{r}_i,q_{n_2},\cdots,q_1)U(z_2,k_2,s_2) .
\end{equation}
\end{widetext}
%Hence, if one unified represents Feynman amplitude
%\begin{eqnarray}
%M=\mu(n_1,s_1)^{+}\gamma_0 \tilde{A_n}\mu(n_2,s_2)
%\begin{cases}
%\mu(n_1,s_1)^{+}\gamma_0(\widehat{p}_{a}\gamma^0-i\gamma^5\widehat{p}_{b}\gamma^0) \mu(n_2,s_2), &\qquad \text{case 1}\\
%\mu(n_1,s_1)^{+}\gamma_0(\widehat{p}_{a}-i\gamma^5\widehat{p}_{b} ) \mu(n_2,s_2), &\qquad \text{case 2}\\
%i\mu(n_1,s_1)^{+}\gamma_0 (\widehat{p}_{b}\gamma^0+i\gamma^5\widehat{p}_{a}\gamma^0)\mu(n_2,s_2), &\qquad \text{case 3} \\
%i\mu(n_1,s_1)^{+}\gamma_0(\widehat{p}_{b}+i\gamma^5\widehat{p}_{a}) \mu(n_2,s_2), &\qquad \text{case 4} 
%\end{cases}
%\label{p19}
%\end{eqnarray}
%there are totally four types of $\tilde{A_n}$, and each case of Eq. (\ref{p17}) (\ref{p18}) is one type, which is successively called case 1, case 2, case 3 and case 4 in the following for convenience.

\subsection{Calculation of scattering amplitudes}
The scattering amplitude of a process with $n$ fermion lines usually takes the form of
\begin{equation}
{\cal M}\sim \sum_i F_{i_1}F_{i_2}\cdots F_{i_n},
\label{eqn:m}
\end{equation}
where the coefficients have been dropped. It is natural to assume that the summation over Lorentz indices is already been done inside each fermion line. After that, we can separate the fermion lines into three types: 1) with no Lorentz indices; 2) with one Lorentz index; 3) with two or more Lorentz indices. In order to calculate the amplitude, we need to calculate all these three types of fermion lines. 

We have already presented a detailed introduction on how to calculation fermion lines without indices, hence there is no need to discuss them here. Meanwhile, as aforementioned fermion lines with one index are Lorentz vectors, whose components can be obtained by introducing four auxiliary vectors. It should be emphasized that once obtained, these vectors can also be contracted into other fermion lines.

So far we have not mentioned how to deal with fermions lines with two or more indices. Of course, they can be taken as Lorentz tensors and all of their components can be obtained with the four auxiliary vectors, just like what is done to those with one index. But there is a better way.

First we would like to point out: at tree level, if there are $n$ pair of Lorentz index contractions among $m$ fermion lines (suppose each fermion line carries at least one index, since those without Lorentz indices are irrelevant), there should be at least one fermion line with just one index. 
The proof of this statement is straightforward: 
\begin{itemize}
\item Index contraction between two fermion lines can be taken as the connection of the two lines by an internal line.
\item At tree level, there can be at most $m-1$ such internal lines, hence $n\le m-1$.
\item If each fermion line carries at least two indices, we will have $2m\le 2n$, which is not allowed at tree level.
\end{itemize}
Based on this, we will never need to calculate fermion lines with two or more indices if we do the calculation in the following way:
\begin{enumerate}
\item separate the fermion lines into three groups as described before.
\item calculate all the fermion lines without indices.
\item calculate a fermion line with one index, take it as a vector and contract it into another lines.
\item back to the top until all the fermion lines are calculated.
\end{enumerate}
Of course there could be some further improvements in performing this such as the order of the lines calculated, but we will not discuss them here.

In the end of this section, we introduce another trick used in the calculation of the amplitude. If a fermion line $F_1^\mu$ is connected with another fermion line $F_2^{\nu\cdots}$ via a massive vector boson whose propagator is $g_{\mu\nu}-{p_\mu p_\nu}/{m^2}$ (other parts dropped), a new vector $k\equiv F_1-({F_1\cdot p}/{m^2})p$ is introduced. Therefore the index contraction becomes
\begin{equation}
F_1^\mu F_2^{\nu\cdots}(g_{\mu\nu}-\dfrac{p_\mu p_\nu}{m^2})=k_\nu F_2^{\nu\cdots} .
\end{equation}
This trick prevents the two terms in the propagator from being separated, hence reduces the total number of terms in the summation of Eq.~(\ref{eqn:m}).

\section{Comparison with other packages\label{sec:compare}}
We have achieved this new algorithm in our FDC package~\cite{Wang:2004du}. At first, we checked its efficiency with the old version in FDC on the calculation of multiple-fermion-line amplitudes and find much improvement.
Second, we check its efficiency on same calculation by using both FDC and some other packages, and present the comparision of the time used 
by different packages in the following. 
%In this section, we show ,  we compare it with some other packages.

\subsection{Comparison with MadGraph}
The first package we have compared with is MadGraph~\cite{Alwall:2014hca}, which might be the most famous package for automatic calculation.
The version of MadGraph used is MadGraph5$\_$aMC@NLO. Since the calculation for a $2\rightarrow2$ process is too fast we choose a $2\rightarrow4$ process $e^+e^-\rightarrow b\bar{b}c\bar{c}$ and a $2\rightarrow6$ process $e^+e^-\rightarrow c\bar{c}c\bar{c}c\bar{c}$ as our benchmark processes. In order to show  the advantage of our new algorithm, charm and bottom quarks are kept massive, and all the polarization configurations are summed over. For $e^+e^-\rightarrow b\bar{b}c\bar{c}$, there are 8 Feynman diagrams at the order of $\alpha^2\alpha_s^2$ while for $e^+e^-\rightarrow c\bar{c}c\bar{c}c\bar{c}$, there are 576 diagrams at the order of $\alpha^2\alpha_s^3$. 

The comparison is done with an Intel i3-4150 dual-core processor, while only one core is used. 
Since MadGraph includes many other built-in functions, it is hard to obtain the exact time used in the calculation of amplitudes.
Hence the comparison is done as follows:
\begin{itemize}
\item Set the number of points in each iteration to be 10000 in MadGraph.
\item Get the time for each iteration, which is provided by MadGraph automatically. Since there might be some optimizations in phase space integration, the time for later iteration is thought to be most accurate.
\item By assuming the efficiency of phase space integration is 100\%, the time taken above is regards as estimated time used by MadGraph for 10000 points in phase space integration.
\item Similar thing is done in FDC, but the efficiency of phase space integration is replaced with the actual one. 
\end{itemize}

\begin{table}[!htb]
	\begin{tabular}{|c|*{4}{c|}} 
		\hline
		%Benchmark Process 
		&
		\multicolumn{2}{c|}{$e^{+}e^{-}\rightarrow b\bar{b}c\bar{c}$} &  
		\multicolumn{2}{c|}{$e^{+}e^{-}\rightarrow c\bar{c}c\bar{c}c\bar{c}$}   \\ 
		\cline{2-3}\cline{4-5}
		\hline
		$\sqrt{s}$(GeV) &  ~~~~FDC~~~~ &  MadGraph & ~~~~FDC~~~~  & MadGraph  \\
		\hline
		20 & 0.6 & 2.26 & 103.4 & 4008  \\
		\hline
		50 & 0.5  & 2.28 &  90.4    & 3936   \\
		\hline
		100 & 0.5 & 2.23 & 111.4 & 3990 \\
		\hline
		200 & 0.5 & 2.24 & 127.9  &  4044 \\
		\hline
		500 & 0.5 & 2.24 & 154.2 & 4002  \\
		\hline
		1000 & 0.5 & 2.23 & 172.8 &  4002  \\
		\hline				
	\end{tabular}
	\caption[]{Estimated time for 10000 points in phase space integration in unit of second.}	
	\label{tab:two}
\end{table}
The comparison is done with several different center-of-mass (c.m.) energies and the results are listed in Table~\ref{tab:two}.
It can be seen from the table that in the calculation of the $2\rightarrow 4$ process, FDC is at least 3 times faster than MadGraph, while in the calculation of the $2\rightarrow 6$ process, FDC is $20\sim40$ times faster than MadGraph as the c.m. energy changes.

\subsection{Comparison with WHIZARD}
Another package we have compared with is WHIZARD~\cite{Kilian:2007gr,Moretti:2001zz}, and the version we use is 2.8.4.

\begin{figure}[hbtp!]
\centering
\includegraphics[width=0.5\textwidth]{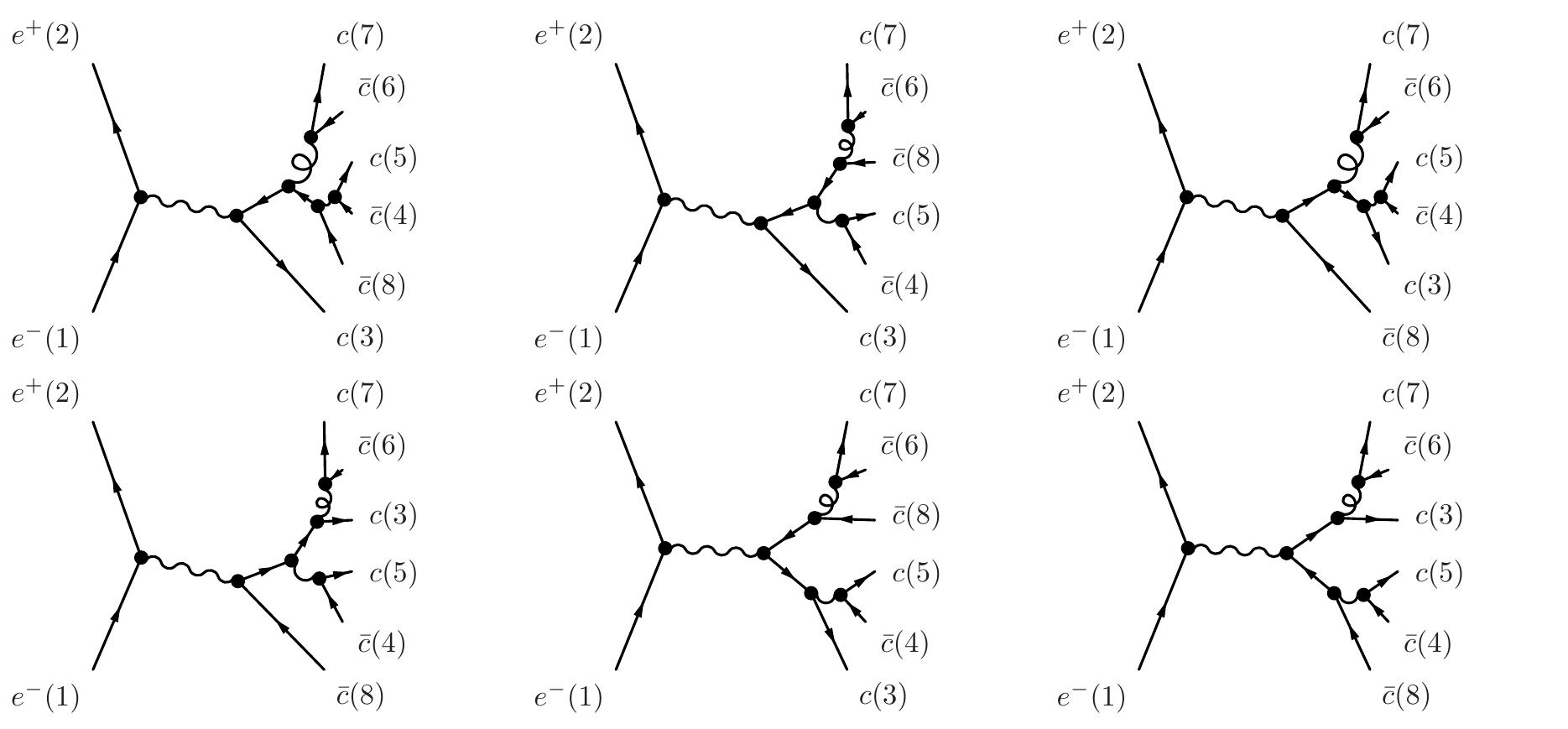}
 \caption{Selected Feynman diagrams of $e^+e^-\rightarrow c\bar{c}c\bar{c}c\bar{c}$ in the comparison with WHIZARD. The vector boson coupled with $e^+e^-$ can be photon or $Z$ boson.}
  \label{fig:whizard}
\end{figure}

Since WHIZARD is unable to give results at certain order of $\alpha$ and $\alpha_s$, we have to choose part of the Feynman diagrams in each processes for comparison. Hence the benchmark processes are changed into:
\begin{itemize}
\item 4 Feynman diagrams of $e^+e^-\rightarrow b\bar{b}c\bar{c}$, in which $b\bar{b}$ is produced via a gluon.
\item Another 4 Feynman diagrams of $e^+e^-\rightarrow b\bar{b}c\bar{c}$, in which $c\bar{c}$ is produced via a gluon.
\item 12 Feynman diagrams of $e^+e^-\rightarrow c\bar{c}c\bar{c}c\bar{c}$, as shown in Fig.~\ref{fig:whizard}.
\end{itemize}

The comparison is done similarly, with the results shown in Table~\ref{tab:one}. Since both FDC and WHIZARD can provide expected time for a certain number of events directly, the time for 10000 events are used this time. Meanwhile, the first two processes are marked with $g^*\rightarrow b\bar{b}$ and $g^*\rightarrow c\bar{c}$, respectively. 
\begin{table*}[!htb]
	\begin{tabular}{|c|*{6}{c|}} 
		\hline
		&
		\multicolumn{2}{c|}{$e^{+}e^{-}\rightarrow b\bar{b}c\bar{c}$ ($g^*\rightarrow b\bar{b}$)} &  
		\multicolumn{2}{c|}{$e^{+}e^{-}\rightarrow b\bar{b}c\bar{c}$ ($g^*\rightarrow c\bar{c}$)} & 
		\multicolumn{2}{c|}{$e^+e^-\rightarrow c\bar{c}c\bar{c}c\bar{c}$}  \\ 
		\cline{2-3}\cline{4-5}\cline{6-7}
		\hline
		$\sqrt{s}$ (GeV)
		& \quad \quad FDC \quad \quad & WHIZARD & \quad \quad FDC \quad \quad & WHIZARD & \qquad \quad FDC \qquad \quad & WHIZARD \\
		\hline
		20 & 0.3 & 4 & 0.3 & 5 & 7.1 & 319\\
		\hline
		50 &  0.3   & 5  &  0.3  &  7 & 7.0 & 760 \\
		\hline
		100 & 0.3 & 6 & 0.3 & 6 & 6.8 & 3935 \\
		\hline
		200 & 0.2 & 8 & 0.3 & 9 & 6.8 & 6999 \\
		\hline
		500 & 0.2 & 8 & 0.3 & 9 & 7.1 &  7905 \\
		\hline
		1000 & 0.2 & 11 & 0.4 & 10 & 7.0 & 6277 \\
		\hline		

	\end{tabular}
	\caption[]{Expected time to generate 10000 events in unit of second. }
	\label{tab:one}	
\end{table*}

It can be seen from the table that in the $2\rightarrow 4$ process, FDC is more than 13 times faster than WHIZARD, while in the $2\rightarrow 6$ process, FDC is dozens to hundreds of times faster. 

%\textcolor{red}{The comparison of pure amplitude calculation is also carried out between WHIZARD and FDC, and the time consuming is almost the same in this condition.}
{\it Note added}: After this work is submitted, a direct comparison on the computation of pure amplitude square with WHIZARD is available, with the help from its authors~\cite{compare}. Little difference is found in the time cost of both packages for this part. The large difference observed in Table~\ref{tab:one} should arise from other parts, such as phase space integrations and event generations.

From both comparisons, it can be concluded that with the new algorithm, FDC gives a very good performance in the processes with multiple fermion lines. 

\section{summary\label{sec:summary}}
In this paper, a new approach for the numerical calculation of fermion lines at tree level is introduced. By calculating two vectors recursively without any matrix multiplications, the result of a fermion line is reduced to a very compact form which depends only on these two vectors. Furthermore, the results for all possible polarization configurations can be obtained at the same time without extra cost. 
As shown in the comparisons, FDC gives a very good performance in the processes of multiple fermion lines  with this new approach and some other improvements. A further comparison with WHIZARD shows that this new approach has a competitive efficiency in computing pure amplitude square without phase space 
integration.

%{\color{red}paused here} 
%
%
%
%
%Some new methods in reduction and calculation of the Feynman amplitude are given. These new methods are mainly focusing on the reduction of multi fermion line and include many parts. A new method to simplify spinor is given and through this skill, the split of the spinor can be avoided. For a Dirac matrices product, a new method is introduced and this product can be uniformly expressed as a recursive graceful form no matter how long this product is. Some other skills are also introduced in the calculation of amplitude. After implementing the above methods in our program FDC, we compare it with other programs respect to time consumption for different processes. Through the comparison, one can see the new method surely greatly improved the calculation speed and can expect that the more final states, the better improvement.
 \section*{Acknowledgements}
We would like to thank Wolfgang Kilian, Thorsten Ohl and J\"urgen Reuter for their help in the comparison with WHIZARD. 
This work was supported by the National Natural Science Foundation of China with Grant Nos. 12135013 and 11975242. It was also supported in part by National Key Research and Development Program of China under Contract No. 2020YFA0406400.
\bibliography{ref.bib}

\end{document}